\documentclass[preprint,showpacs,onecolumn,nofootinbib]{revtex4}
\pdfoutput=1
\usepackage[colorlinks=true,linkcolor=blue,urlcolor=blue,filecolor=black,citecolor=red,pdfstartview=FitV,pdftitle={},pdfsubject={},pdfkeywords={},pdfpagemode=None,bookmarksopen=true]{hyperref}
\usepackage{graphicx}
\usepackage{amsmath}
\usepackage{amsfonts}
\usepackage{amssymb,ulem}
\usepackage{color}%
\usepackage{dcolumn}
\usepackage{MnSymbol,wasysym}
\usepackage{braket}
\usepackage{eurosym}
\usepackage{calrsfs}
\usepackage[usenames,dvipsnames,svgnames]{xcolor}
\usepackage{color}%

\begin{document}

\baselineskip=0.6 cm
\title{Scalar Field in Massive BTZ Black Hole and Entanglement Entropy}
\author{Yu-Ting Zhou$ ^{1,2}$}\email{constaantine@163.com}
\author{Xiao-Mei Kuang$ ^{2}$}\email{xmeikuang@yzu.edu.cn}
\affiliation{$^1$ College of Mathematics and Science, Yangzhou University, Yangzhou 225009, China}
\affiliation{$^2$ Center for Gravitation and Cosmology, College of Physical Science and Technology,
Yangzhou University, Yangzhou 225009, China}
\date{\today }

\vspace*{0.2cm}
\begin{abstract}
\baselineskip=0.6 cm
In this paper, we investigate the quantum scalar fields in massive BTZ black hole background. We study the entropy of the system by evaluating the  entanglement entropy with the use of discretized approach. Specifically, we fit the results with $\log$ -modified formula of the  black hole entropy which is introduced by quantum correction. The coefficients of leading and sub-leading terms affected by the mass of graviton are numerically analyzed.
\end{abstract}

\maketitle
\newpage

\section{Introduction}\label{sec:Introduction}
Black holes are solutions of Einstein equation in general gravitational theories, and the black hole is one of   fascinating parts of our universe.  Black holes are thermal systems because they were found to have temperature and entropy. The temperature of a black hole is proportional to  the surface gravity at the event horizon. In the case of  general relativity, the entropy, known as
Bekenstein-Hawking entropy, is usually proportional to the area of the event horizon in the framework of black hole thermodynamics \cite{Israel:1967wq,Unruh:1976db,Hawking,Shankaranarayanan:2000gb,Angheben:2005rm,Kerner:2006vu}\footnote{It is noticed that beyond general relativity, the entropy is not simply  proportional to the surface area of the horizon and additional terms would appear, see for example \cite{Clunan:2004tb,Jacobson:1993xs,Chakraborty:2015wma}.}.
However, when one considers quantum fields in the vicinity of black holes, the area law of entropy will get an additional quantum corrections. Till now, physicists have made plenty of  attempts to disclose  the microscopic essence of the black hole entropy and its connection with the area of event horizon.   For instance,  the authors of \cite{Gibbons,Brown,Hawking2} computed the entropy by evaluating the Euclidean action. The connecting between  the entropy and  instanton amplitude, which describes  the pair production of charged  black holes, can be seen in \cite{Garfinkle}.  The proposal that the entropy is the Noether charge of the bifurcate Killing horizon  has been addressed in \cite{Wald,Iyer}.  Later, the symmetry based approach to black hole entropy has been proposed, which connects  the central charge of conformal field theory to the black hole entropy\cite{Strominger,Carlip2}. This proposal of the central charge of the Virasoro algebra on the horizon has been  generalized into the case of  Witt algebra\cite{Dreyer:2013noa}, surface charge algebra\cite{Barnich:2007bf}, Virasoro algebra null surfaces\cite{Chakraborty:2016dwb}  and  $2-$cocycles on the Lie algebra\cite{Barnich:2001jy} and so on.

We know that the consideration of quantum mechanics gives rise to the thermal Hawking radiation, which does not carry information.   One viewpoint is that the quantum gravity theory, such as string theory and loop quantum gravity, is called for to understand this information loss paradox.  However, it was addressed in \cite{Mathur:2009hf} that the interior of black holes has a `fuzzball' structure. This  gives a qualitative picture of how classical idea  breaks down in black hole physics and the information paradox can be resolved. Especially, the authors somehow took care of   both the near horizon behaviors and possible existence of singularity instead of focusing on one of them in the discussion.
It is noticed that the microscopic derivation of black hole entropy was proposed with D-brane method in string theory\cite{Polchinski} and in brick wall model\cite{tHooft}. On the other hand, the entanglement entropy is a measure of the correlation between subsystems, which are separated by a boundary called the entangling surface. It is a measure of the information loss due to the division of the system, and so it depends on the geometry of the boundary \cite{Bombelli,Srednicki}. The entanglement entropy model, as one of the most attractive candidate of the black hole entropy, has been  widely studied\cite{Frolov,Callan,Kabat,Holzhey,Mukohyama,Cadoni:2007vf,Cadoni:2009tk,Hung:2011ta}.

In this paper, we aim to investigate the entropy of massive BTZ black hole\cite{Hendi:2016pvx,Chougule:2018cny} via calculating its entanglement entropy by introducing
a massless quantum scalar field. We shall apply the discretized approach to evaluate the entropy of the quantum scalar field in the spacetime by analyzing the similar harmonic oscillator, and we closely follow the procedure shown in\cite{Bombelli,Shiba:2013jja,Shiba:2012np}.
We note that, similar studies on (rotating) BTZ black hole with massless or massive scalar field have been done in \cite{Singh:2011gd,Sachan:2014hna}, where the authors numerically analyzed the effect of angle momentum $(J)$ and mass of scalar field $(m_s)$ on the coefficients in the entanglement entropy of BTZ black hole.

 Here, we shall focus on the effect of the mass of graviton on the entanglement  entropy  in three dimensional massive black hole.
As is known that black holes provide a particle environment for testing gravity and are incredibly important theoretical tools for exploring general gravity, and Schwarzschild black hole is the most general spherically symmetric vacuum solution. Massive gravity  theory is one of the theories  beyond Einstein's gravity theory with  massless graviton. In recent years, some cosmologists propose the idea of massive graviton to modify general relativity. One can also make contributions to explain the accelerated expansion of the universe without dark energy\cite{DAmico:2011eto}.  Early attempt of the construction in massive gravity has been established, such as  linear theory of gravity\cite{Fierz:1939ix} and nonlinear model for massive gravity \cite{Vainshtein:1972sx}. Recently, significant progress has been made in constructing massive gravity theories, which would avoid  instability, see for example \cite{deRham:2010ik,deRham:2010kj,Hassan:2011hr,Hassan:2011vm,Hassan:2011tf,Desai:2010ea,Ludeling:2012cu}. Various phenomenology of massive gravity has also been investigated widely.
Moreover, the massive terms in the gravitational action break the diffeomorphism symmetry in the bulk, which corresponds to momentum dissipation in the dual boundary field theory\cite{Blake:2013bqa}.

This paper is organized as follows. We briefly review the massive BTZ black hole in three dimensional gravity theory in section \ref{sec-massiveBTZ}. We study the formula of the entanglement entropy of the massless quantum scalar field, and then evaluate the (sub-)leading coefficients of the entropy by fitting the numerical results in section \ref{sec-EEofScalar}. Section \ref{sec-conclu} contributes to our conclusions.

\section{Massive neutral BTZ black hole}\label{sec-massiveBTZ}
We first briefly review the  three dimensional  Einstein-massive gravity with the action \cite{Hendi:2016pvx,Chougule:2018cny}
\begin{equation}
\mathcal{I}=-\frac{1}{16\pi }\int d^{3}x\sqrt{-g}\left[ \mathcal{R}-2\Lambda+m^{2}\sum_{i}^{}c_{i}\mathcal{U}_{i}(g,f)\right] ,
\label{Action}
\end{equation}%
 where $\mathcal{R}$ is the scalar curvature, $\Lambda=-1/L^2$ is the
cosmological constant, and the last terms are Fierz-Pauli mass terms  where $m$ denotes  the mass of graviton\cite{Creminelli:2005qk}. Here,  $c_{i}$ are constants and $\mathcal{U}_{i}$'s are symmetric polynomials of the
eigenvalues of the $3\times 3$ matrix $\mathcal{K}_{\nu }^{\mu }\equiv\sqrt{%
g^{\mu \alpha }f_{\alpha \nu }}$. In this matrix,  $g$ denotes the background metric, and  the fixed rank$-2$ symmetric
tensor $f$ is  the reference metric  which is  inevitably included to construct the massive term of graviton in massive gravity and its form will be chosen later.  As addressed in \cite{deRham:2010kj,Hinterbichler:2011tt}, $\mathcal{U}_{i}$ can be written as
\begin{eqnarray}\label{eq-Ui}
\mathcal{U}_{1} &=&\left[ \mathcal{K}\right] ,\;\;\;\;\;\mathcal{U}_{2}=%
\left[ \mathcal{K}\right] ^{2}-\left[ \mathcal{K}^{2}\right] ,\;\;\;\;\;%
\mathcal{U}_{3}=\left[ \mathcal{K}\right] ^{3}-3\left[ \mathcal{K}\right] %
\left[ \mathcal{K}^{2}\right] +2\left[ \mathcal{K}^{3}\right] ,  \notag \\
&&\mathcal{U}_{4}=\left[ \mathcal{K}\right] ^{4}-6\left[ \mathcal{K}^{2}%
\right] \left[ \mathcal{K}\right] ^{2}+8\left[ \mathcal{K}^{3}\right] \left[
\mathcal{K}\right] +3\left[ \mathcal{K}^{2}\right] ^{2}-6\left[ \mathcal{K}%
^{4}\right],\cdots
\end{eqnarray}
where the dots denote the higher order terms of $\mathcal{K}$. The square root in
$\mathcal{K}$ denotes matrix square root, i.e, $(\sqrt{\mathcal{K}})^{\mu}_{\nu}(\sqrt{\mathcal{K}})^{\nu}_{\kappa}=\mathcal{K}^{\mu}_{~~\kappa}$, and the rectangular brackets denotes traces, i.e., $[\mathcal{K}]\equal\mathcal{K}^{\mu}_{\mu}$ and $[\mathcal{K}^2]\equal(\mathcal{K}^2)^{\mu}_{~~\mu}$ with $(\mathcal{K}^2)_{\mu\nu}\equal\mathcal{K}_{\mu\alpha}\mathcal{K}^{\alpha}_{~~\nu}$. We note that the study of ghost-free massive gravity can be seen in \cite{Hassan:2011hr,Hassan:2011tf}.

The action (\ref{Action}) gives us the Einstein equation
\begin{equation}
G_{\mu \nu }+\Lambda g_{\mu \nu }+m^{2}\chi _{\mu \nu }=0,
\label{Field equation}
\end{equation}%
where the Einstein tensor $G_{\mu \nu }$ and massive term $\chi _{\mu \nu }$ are
\begin{eqnarray}
G_{\mu \nu }&=&R_{\mu \nu }-\frac{1}{2}g_{\mu \nu }R,\\
\chi _{\mu \nu } &=&-\frac{c_{1}}{2}\left( \mathcal{U}_{1}g_{\mu \nu }-%
\mathcal{K}_{\mu \nu }\right) -\frac{c_{2}}{2}\left( \mathcal{U}_{2}g_{\mu
\nu }-2\mathcal{U}_{1}\mathcal{K}_{\mu \nu }+2\mathcal{K}_{\mu \nu
}^{2}\right) -\frac{c_{3}}{2}(\mathcal{U}_{3}g_{\mu \nu }-3\mathcal{U}_{2}%
\mathcal{K}_{\mu \nu }+  \notag \\
&&6\mathcal{U}_{1}\mathcal{K}_{\mu \nu }^{2}-6\mathcal{K}_{\mu \nu }^{3})-%
\frac{c_{4}}{2}(\mathcal{U}_{4}g_{\mu \nu }-4\mathcal{U}_{3}\mathcal{K}_{\mu
\nu }+12\mathcal{U}_{2}\mathcal{K}_{\mu \nu }^{2}-24\mathcal{U}_{1}\mathcal{K%
}_{\mu \nu }^{3}+24\mathcal{K}_{\mu \nu }^{4}).  \label{massiveTerm}
\end{eqnarray}
In order to solve the equation of motion, we take the ansatz of the metric as
\begin{equation}
ds^{2}=-f(r)dt^{2}+f^{-1}(r)dr^{2}+r^{2}d\varphi ^{2},  \label{metric}
\end{equation}
where $f(r)$ is an arbitrary function of radial coordinate, and we will work with $L=1$ so that $\Lambda=-1$.  Also following \cite{Vegh:2013sk}, we choose the ansatz of the
reference metric as \footnote{In principle, the choice of the reference metric could be arbitrary\cite{Hassan:2011tf}, however, here we follow \cite{Vegh:2013sk}
to choose the form  \eqref{f11}. As is addressed in \cite{Vegh:2013sk}, with this choice, the graviton mass term preserves general covariance in the $t$ and $r$ coordinates, but breaks it in spatial coordinates. Then only one field $\phi^{\varphi}$ is needed due to the spatial reference metric,  so that the bulk could be seen to fill with homogeneous solid. Moreover, this makes us manage to find the analytical solution of the black hole. }
\begin{equation}
f_{\mu \nu }=\partial_{\mu}\phi^a\partial_{\nu}\phi^b\eta_{ab}=diag(0,0,c^{2}),  \label{f11}
\end{equation}%
where $c$ is a positive constant, $\phi^a(x)$ is kind of coordinate transformation using $\eta_{ab}$ and different choices for the $\phi^a$ fields correspond to different gauges. Then with the metric ansatz (\ref{f11}),  $\mathcal{U}_{i}$'s in \eqref{eq-Ui} can be calculated as\cite{Hassan:2011hr,Cai:2014znn}
\begin{equation}
\mathcal{U}_{1}=\frac{c}{r},\;\;\;\;\;\mathcal{U}_{2}=\mathcal{U}_{3}=%
\mathcal{U}_{4}=\cdots=0,  \label{U}
\end{equation}
which means that  in three dimensional case, the only contribution of massive terms comes from the term $\mathcal{U}_{1}$ in the action.

Subsequently,  the independent Einstein equations are
\begin{eqnarray}
&&rf^{\prime }(r)-2r^{2} -m^{2}cc_{1}r =0,  \label{eqENMax1} \\
&&\frac{r^{2}}{2}f^{\prime \prime }(r)-r^{2} =0,
\label{eqENMax2}
\end{eqnarray}%
the  solution to which is
\begin{equation}
f(r)=r^{2}-M +m^{2}cc_{1}r.
\label{f(r)ENMax}
\end{equation}%
Here $M$ is an integration constant which is related to the total mass
of the black hole. We note  that in the absence of massive term with $m=0$, the solution recovers neutral  BTZ black hole $f\left( r\right) =r^{2}-M$ without rotation.

Applying the standard method,  we  obtain the temperature of the black hole as
\begin{equation}
T=\frac{ r_{+}}{2\pi }+\frac{m^{2}cc_{1}}{%
4\pi },\label{TotalTT}
\end{equation}
where $r_{+}$ is the event horizon satisfying $f(r_+)=0$
and the Hawking entropy of the black hole is
\begin{equation}
S=\frac{\pi }{2}r_{+}.  \label{TotalS}
\end{equation}

For the convenience of further study, we shall do kind of transformation of the metric.
The solutions to $f(r)=0$ are
\begin{equation}
r_{\pm}=\frac{-m^{2}cc_{1}\pm\sqrt{m^{4}c^{2}c_{1}^{2}+4M}}{2}.\label{newhorizon}
\end{equation}
Then, we introduce the proper length, $\rho$, by the coordinate transformation
\begin{equation}
r^{2}=r^{2}_{+}cosh^2\rho+r^{2}_{-}sinh^{2}\rho. \label{properLength}
\end{equation}
Subsequently, the metric of the black hole can be rewritten in terms of  the proper length as
\begin{equation}
ds^{2}=-u^{2}dt^{2}+d\rho^{2}+\frac{[m^{2}+\sqrt{m^{4}+4(M+u^{2})}]^{2}}{4}d\varphi^{2}\label{properLmetric}
\end{equation}
where we have defined $u^{2}=r^{2}-M+m^{2}cc_{1}r$. In the following study, we will set $c=c_1=1$ without loss of generality.

\section{Entanglement entropy of massive BTZ black hole with scalar fields}\label{sec-EEofScalar}

\subsection{Free massless scalar fields in massive BTZ black hole}
The action of massless scalar field in the curved spacetime is
\begin{equation}
S_{scalar}=-\frac{1}{2}\int d^3x\sqrt{-g}g^{\mu\nu}\partial_{\mu}\Phi\partial_{\nu}\Phi.\label{MSFaction}
\end{equation}
In the background of massive BTZ black hole \eqref{properLmetric}, considering the cylindrical symmetry of the system and the form of the scalar field
\begin{equation}
\Phi(t,\rho, \varphi)=\sum_{n}\Phi_{n}(t,\rho)e^{in\varphi},\label{wavefunction}
\end{equation}
we then further evaluate the action \eqref{MSFaction} as
\begin{eqnarray}
S_{scalar}=-\frac{1}{2}\int d^3x&&\sum_{n}\Big(-\frac{m^{2}+\sqrt{m^{4}+4(M+u^{2})}}{2u}\dot{\Phi}_{n}^{2}+\frac{u[m^{2}+\sqrt{m^{4}+4(M+u^{2})}]}{2}(\partial_{\rho}\Phi_{n})^{2}
 \nonumber \\&&+n^{2} \frac{2u}{m^{2}+\sqrt{m^{4}+4(M+u^{2})}}\Phi_{n}^{2}\Big)=\int d^3x\sum_{n}\mathcal{L}_n ,\label{massiveBHaction}
\end{eqnarray}
where we defined  $\mathcal{L}_n $ as the Lagrangian density for the $n-$mode.

The conjugate momentum corresponding to $\Phi_{n}$ is given by
\begin{equation}
\pi_{n}=\frac{\delta\mathcal{L}_n}{\delta \dot\Phi_n}=\frac{2un^{2}}{m^{2}+\sqrt{m^{4}+4(M+u^{2})}}\Phi_{n}. \label{conjmomentum}
\end{equation}
Therefore, the Hamiltonian density of the system is
\begin{equation}
H=\sum_{n}H_n=\sum_{n}\left(\frac{1}{2}\int d\rho  \pi_{n}^{2}(\rho)+\frac{1}{2}\int d\rho d\rho' \psi_{n}(\rho)V_{n}(\rho,\rho')\psi_{n}(\rho')\right) \label{Hamiltonian}
\end{equation}
where we introduced
\begin{equation}
\psi_{n}(t,\rho)=\sqrt{\frac{2u}{m^{2}+\sqrt{m^{4}+4(M+u^{2})}}}\Phi_{n}(t,\rho) \label{redefinephi}
\end{equation}
and
\begin{eqnarray}\label{protentailterm}
\psi_{n}(\rho)V_{n}(\rho,\rho')\psi_{n}(\rho')=&& \frac{u[m^2+\sqrt{m^{4}+4(M+u^{2})}]}{2}\Big(\partial_{\rho}(\sqrt{\frac{2u}{m^{2} +\sqrt{m^{4}+4(M+u^{2})}}})\psi_{n}\Big)^{2}\nonumber \\
&&+\frac{2un^{2}}{(m^{2}+\sqrt{m^{4}+4(M+u^{2})})^{2}} \psi_{n}^{2}.
\end{eqnarray}

To proceed, we discretize the system for the convenience of computation via
\begin{equation}
\rho \rightarrow (A-1/2)a,  \qquad   \delta(\rho-\rho') \rightarrow\delta_{AB}/a \label{discrecondition}
\end{equation}
where A,B=1,2 ...N and `a' is the UV cut-off length, so that $N\propto r_+/a$. We note that the continue limit recovers when $a\rightarrow 0$ and $N\rightarrow \infty $ as the size of system is fixed.
Then, the Hamiltonian of the discretized system can be obtained by replacing
\begin{equation}
\psi_{n}(\rho) \rightarrow q^{A},\qquad \pi_{n}(\rho) \rightarrow p_{A}/a,\qquad V(\rho,\rho') \rightarrow V_{AB}/a^{2} \label{Hamiltonreplace}
\end{equation}
in \eqref{Hamiltonian}, the expression of which is then
\begin{equation}\label{HD}
H_D=\sum_{A,B=1}^N\left(\frac{1}{2a}\delta^{AB}p_{nA}p_{nB}+\frac{1}{2}V_{AB}^{n}q_n^A q_n^B\right).
\end{equation}
Here, $ V_{AB}^{n}$ is $N\times N$ matrix representation given as
\begin{eqnarray}
   (V_{AB}^{(n)})=\begin{bmatrix}
    \Sigma_{1}^{(n)} & \Delta_{1} &  &  &  &   \\
     \Delta_{1}  & \Sigma_{2}^{(n)} & \Delta_{2} &  &  &    \\
       & \ddots & \ddots & \ddots & &  \\
       & & \Delta_{A-1} & \Sigma_{A}^{(n)} & \Delta_{A} & & \\
        & & & \ddots &\ddots &\ddots \\
         \end{bmatrix} ,
\end{eqnarray}
with
\begin{eqnarray}\label{sigma}
\Sigma_{A}^{(n)}=&&\frac{2u_{A}}{m^{2}+\sqrt{m^{4}+4(M+u_{A}^{2})}}     \Big(\frac{u_{A+1/2}[m^{2}+\sqrt{m^{4}+4(M+u_{A+1/2}^{2})}]}{2}  \nonumber\\ &&-   \frac{u_{A-1/2}[m^{2}+\sqrt{m^{4}+4(M+u_{A-1/2}^{2})}]}{2}          \Big )
+n^{2}\frac{4u^{2}}{[m^{2}+\sqrt{m^{4}+4(M+u_{A}^{2})}]^{2}}
\end{eqnarray}
and
\begin{eqnarray}\label{delta}
\Delta_{A}=&&-\frac{u_{A+1/2}[m^{2}+\sqrt{m^{4}+4(M+u_{A+1/2}^{2})}]}{2} \sqrt{\frac{2u_{A+1}}{m^{2}+\sqrt{m^{4}+4(M+u_{A}^{2})}}}\nonumber\\ &&\times\sqrt{\frac{2u_{A}}{m^{2}+\sqrt{m^{4}+4(M+u_{A+1/2}^{2})}}}.
\end{eqnarray}
With the Hamiltonian \eqref{HD}-\eqref{delta} in hands, we are ready to apply the method shown in appendix \ref{appendix} to evaluate the entanglement entropy
of the system.

\subsection{Numerical result of entanglement entropy}
Following appendix \ref{appendix}, the entanglement entropy of the above system is given by\cite{Huerta:2011qi}
\begin{equation}\label{entanglemententropy}
S=\lim\limits_{N\to\infty}S(n_{B},N)=S_{0}+2\sum_{n=1}^{\infty}S_{n},
\end{equation}
where $S(n_{B},N)$ is the entanglement entropy of the total system $N$ with partition $n_{B}$; $S_{0}$ is the entanglement entropy of the system for $n=0$ and $S_{n}$ is the entropy of the subsystem for a given `n', respectively.

We first study the entanglement entropy of the massive BTZ black hole at large N, which means that the change of result is not significant with increasing $N$. The numerical results are shown
in table \ref{table1}. We numerically calculate the $S_n$ for
$n_B=10,50$ and $100$ at fixed $N=100$, and then we sum over `n' as $S_{total}$. Specially, we show the results with different mass of graviton. The properties are summarized as follows. First, for fixed $N$ and $m$, larger $n_B$ corresponds to bigger $S_n$ and $S_{total}$. Second, with fixed $n_B$ and $m$, $S_n$ decreases for larger $n$ and finally it becomes slightly significant to the total entanglement entropy. These properties are specifically similar to those observed in \cite{Singh:2011gd}. Third, with fixed $n_B$,
$S_{total}$ is suppressed by stronger mass of graviton.

\begin{table}[h]
\footnotesize
\begin{center}
\begin{tabular}{|c|c|c|c|c|c|c|c|c|c|c|c|c|c|}
 \hline
$n_B$ &\multicolumn{3}{|c|}{$S_0$} & \multicolumn{3}{|c|} {$S_1 $} & \multicolumn{3}{|c|} {$S_2 $} &$\cdots$ &\multicolumn{3}{|c|} {$S_{total} =S$} \\  \hline
$$ &m=0 &m=0.5 & m=1 & m=0 & m=0.5 & m=1 &m=0 &m=0.5&m=1  &$\cdots$ &m=0 &m=0.5&m=1  \\  \hline
$10$&0.98177 &0.98190 &0.98145  &0.00732 &0.00548 &0.00331 &0.00010&0.00046  &0.00027 &$\cdots$ &0.99795  &0.99399 &0.98875  \\  \hline
$50$&1.70383 &1.70259 &1.68877 &0.25020  &0.18895 &0.15193 &0.02178 &0.01702&0.013599  &$\cdots$ &2.25842  &2.12277 &2.02639  \\  \hline
$100$&2.39708  &2.38928 &2.33553  &0.79353  &0.72815 &0.67098 &0.14988  &0.12739 &0.156145  &$\cdots$ &4.35903  &4.16412 &4.06405    \\  \hline
 \end{tabular}
 \caption{\label{table1}The effect of mass of graviton on the entanglement entropy with fixed $N=100$ for different $n_B$.}
 \end{center}
\end{table}

\begin{figure}[htbp]
\centering
\includegraphics[width=0.45\textwidth]{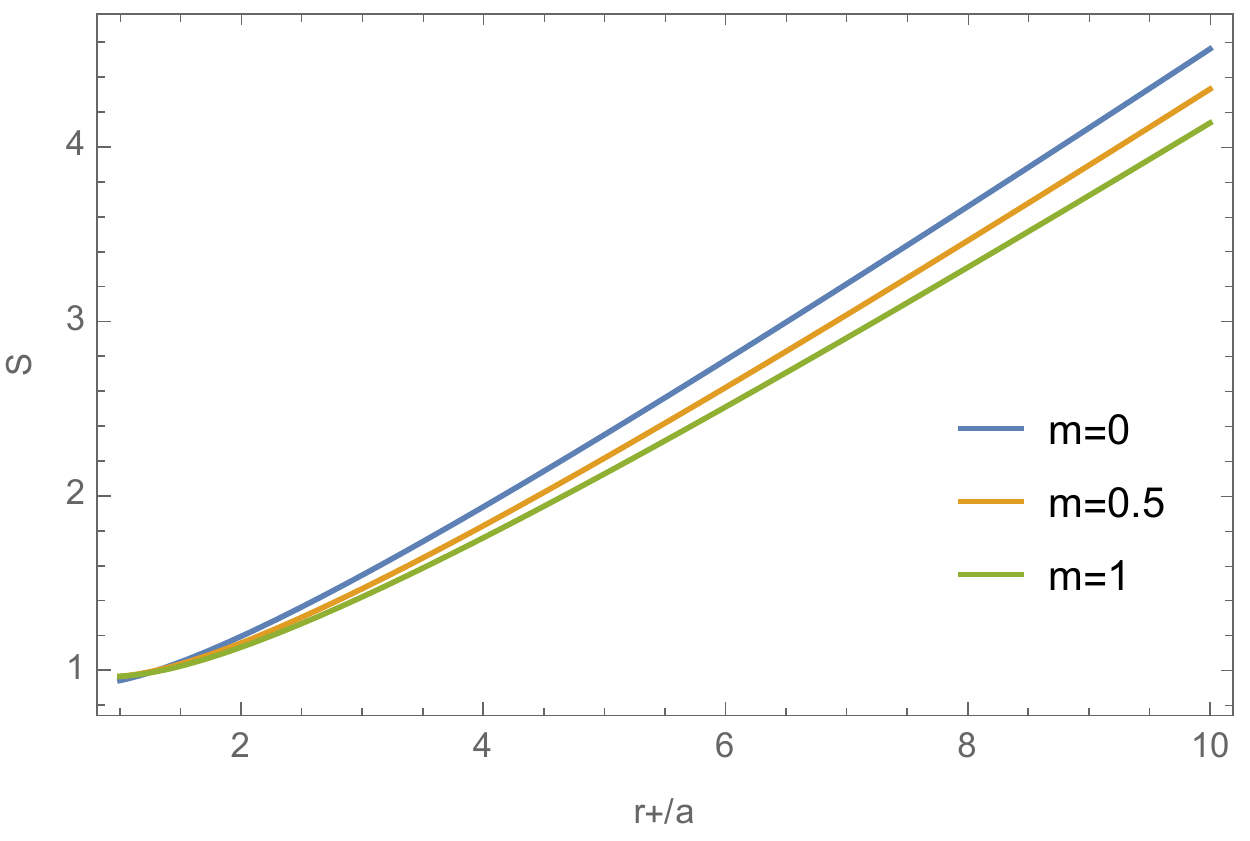}
\caption{The entropy fitted by logarithmic correction \eqref{eq-Slog} for different mass of graviton.}\label{fig-fitlog}
\end{figure}

\begin{figure}[htbp]
\centering
\includegraphics[width=0.45\textwidth]{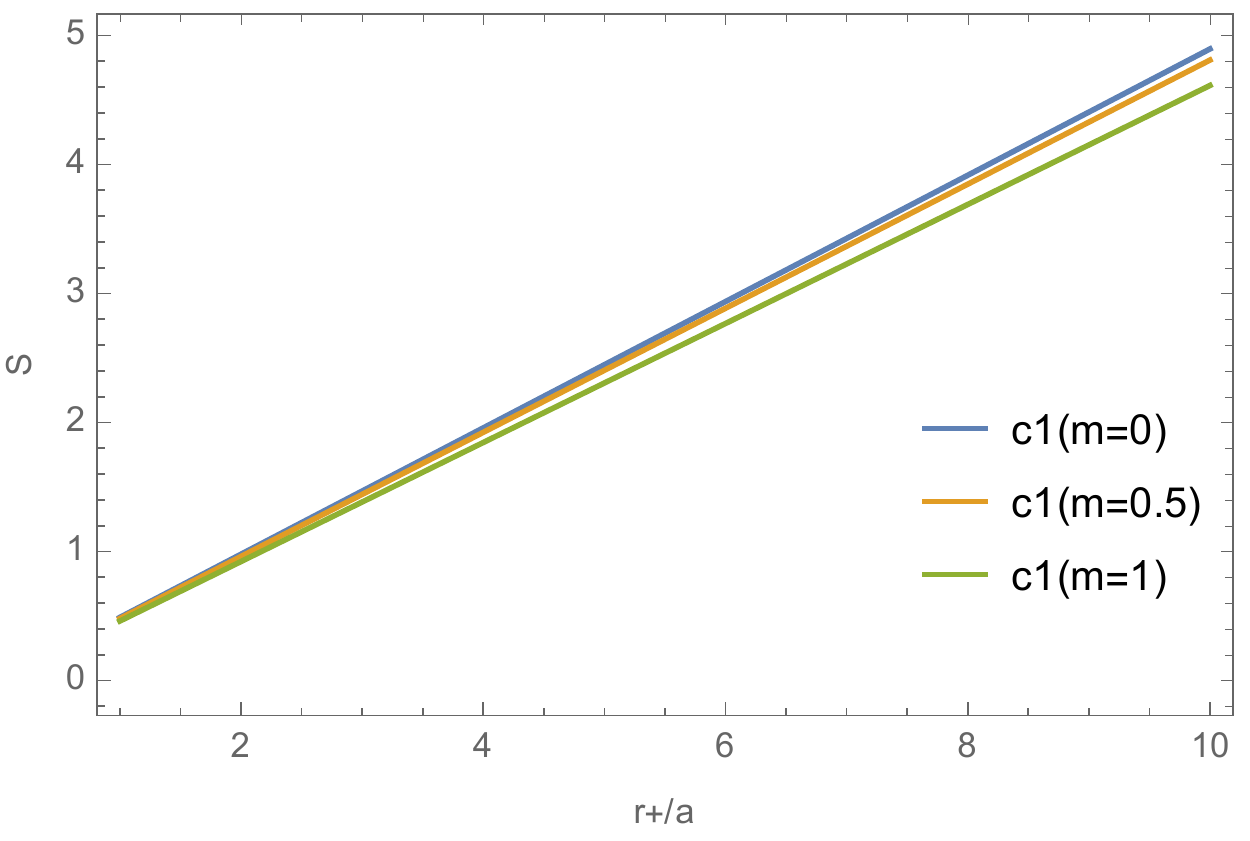}
\caption{Linear fitting of the entropy for different $m$.}\label{fig-fitlinear}
\end{figure}

Next, we numerically evaluate the entanglement entropy of the system with finite $N$. The entropy of the massive BTZ black hole is proportional to the area of the horizon as shown in \eqref{TotalS}, which we rewrite as $S= c_1 r_+/a$ and $c_1$ is a constant while $a$ is the UV cut-off used to discretize the system. Considering the quantum effect, we should estimate the logarithmic correction to the black hole entropy as\footnote{The entanglement entropy we study here is defined by the von Neumann entropy relation
$S_A=-Tr_A (\rho_{A}\log\rho_{A})$ of the system with black hole and the scalar field. Whether this entanglement entropy is exact the black hole entropy \eqref{TotalS} is still an open question deserving further study. We choose the fitting scheme because usually any entropy satisfies  the area law  and involves log correction from quantum affect. It is noticed that the sub-sub-leading corrections may be flexible, but the choice could not affect the qualitative results of the (sub)-leading terms. }
\begin{equation}\label{eq-Slog}
S_q=c_1 r_+/a+c_2\log(r_+/a)+c_3
\end{equation}
where $c_1, c_2$ and $c_3$ are all coefficients to be determined. We numerically calculate the entropy \eqref{entanglemententropy} and fit the results by \eqref{eq-Slog} shown in figure \ref{fig-fitlog}. The fitted coefficients affected by the mass of graviton are shown in table \ref{table2}.
We see that the coefficient of the linear term, $c_1$, decreases as $m$, which is further fitted in figure \ref{fig-fitlinear}.   $c_1$ in the entanglement entropy is dependent of $m$, which is different from the constant in the Hawking entropy \eqref{TotalS}. This is reasonable because as we mentioned in footnote 3 that whether this entanglement entropy is exactly the black hole entropy is still an open question. It is worthwhile to point out that the entanglement entropy of massive black hole has also holographically studied in\cite{Zeng:2015tfj,Zhou:2019jlh} via Ryu-Takayanagi formula\cite{Ryu:2006bv}, and it was found that the massive of graviton would affect the entanglement entropy. So we argue that the massive term has explicit print on the entanglement entropy of black hole  could be a universal property in massive gravity, even though the deep physics call for further study.  Our result shows that the effect of mass of graviton on the entropy is similar to that of mass of scalar field observed in \cite{Sachan:2014hna}. The sub-leading coefficient, $c_2$ is minus and also decreases as the mass  increases while the fitting of $c_3$ is positive and bigger in massive BTZ black hole.
\begin{table}[h]
\begin{center}
\begin{tabular}{|c|c|c|c|}
 \hline
$c$ &{$m=0$}&{$m=0.5$} & {$m=1 $} \\  \hline

$c_1$&0.489594   &0.480959  &0.461284   \\  \hline
$c_2$&-0.34156   &-0.417998   &-0.425697    \\  \hline
$c_3$&0.452134 &0.484263  &0.504732    \\  \hline
 \end{tabular}
 \caption{\label{table2}Coefficients in \eqref{eq-Slog} by fitting the entropy.}
 \end{center}
\end{table}

\section{Conclusion and discussion}\label{sec-conclu}
In this paper, we have evaluated the entanglement entropy of massive BTZ black hole by introducing
a massless quantum scalar field with the use of the discretized approach. We found that in all cases, $S_n$ decreases as  $n$ increases and finally it is not significant to the total entropy. Meanwhile, when we increase $n_B$, the entropy increases. The results are very similar to that found in \cite{Singh:2011gd}. We also obtained that the larger mass of graviton would reduce the entropy. This means that comparing to the theory with massless graviton, the entropy computed by entanglement entropy in massive BTZ black hole is suppressed. Furthermore, we fitted the entropy by the quantum correction formula \eqref{eq-Slog} and found the effect of massive term on the fitting coefficients. The linear coefficient decreases as $m$ increases, which is similar to the effect of the mass of scalar field obtained in \cite{Sachan:2014hna}. The sub-leading coefficient of $\log$ term is suppressed while the constant $c_3$ is enhanced in massive BTZ black hole. Reminding that $m$ also explicitly affects the holographic entanglement entropy in massive gravity,  we argued that it is universal that the massive term has print on the entanglement entropy of black hole in massive gravity, but the deep physics deserve further study.

As it was investigated in \cite{Singh:2014cca}, it would be very interesting to extend our study by introducing Fermion field  instead of scalar field into massive gravity and see the effect of massive term. We shall present the study elsewhere in the near future.

\appendix
\section{Model of entanglement Entropy}\label{appendix}
In this appendix, we review the  discretized method to calculate the entanglement entropy for scalar fields\cite{Bombelli}. Consider the system with coupled harmonic oscillators $q^A,~ (A=1,.......,N)$, the Hamiltonian of which is
\begin{equation}
H=\frac{1}{2a}\delta^{AB}p_Ap_B+\frac{1}{2}V_{AB}q^Aq^B.
\end{equation}
Here $p^A$ and $p^B$ are canonical momentums conjugate to the $q^A$ and $q^B$, respectively, which are defined by the relation $p_A=a\,\delta_{AB}\,\dot{q}^B$ with the Kronecker delta $\delta_{AB}$. $V_{AB}$ is real, symmetric, positive definite matrix and ``a''  is fundamental length characterizing the system. By defining a symmetric, positive definite matrix via $V_{AB}=W_{AC}W^C_B$, one can rewrite the total Hamiltonian as,
\begin{equation}
H=\frac{1}{2a}\delta^{AB}\left(p_A+iW_{AC}q^C\right)\left(p_B-iW_{BD}q^D\right)+\frac{1}{2a}\,\mathrm{Tr}~ W.
\end{equation}
The new operators $(p_{A}+iW_{AC}q^{C})$ and $(p_{B}-iW_{BD}q^{D})$ are the  annihilation  and  creation operators respectively, which are similar to those of  harmonic oscillator problem and they obey similar commutation relation, $[a_A,a_B^{\dagger}]=2W_{AB}$.

In the harmonic oscillator system, then the ground state $\psi_{0}$ satisfies
\begin{equation}
(p_{A}-iW_{AC}q^{C})|\psi_{0}>=0
\end{equation}
and the solution is \cite{Bombelli}
\begin{eqnarray}
\psi_{0}(\{q^C\})&=&<\{q^C\}|\psi_{0}>
=\left[\det\frac{W}{\pi}\right]^{1/4} \mathrm{exp} \left[-\frac{1}{2}W_{AB}\,q^{A}\,q^{B}\right].
\end{eqnarray}
Therefore, the density matrix of the ground state is
\begin{eqnarray}
\rho\left(\{q^A\},\{q^{\prime B}\}\right)&=&<\{q^A\}|0><0|\{q^{\prime B}\}>
=\left[\mathrm{det}\frac{W}{\pi}\right]^{1/2}\mathrm{exp}\left[-\frac{1}{2}W_{AB}\,(q^{A}\,q^{B}+q^{\prime A}\,q^{\prime B})\right].
\end{eqnarray}
We divide ${q^{A}}$ into two subsystems, ${\{q^{a}\}}$ $(a=1,2,.......n_B)$ and $\{q^\alpha\}$ $(\alpha=n_B+1,n_B+2,.......N)$. Then the reduced density matrix of one subsystem is obtained by tracing the degrees of freedom of the other subsystem  as\footnote{The two subsystems are distinguished with label ``$a$'' and  label ``$\alpha$'', respectively.}
\begin{eqnarray}
\rho\Big(\{q^{\prime a}\},\{q^{\prime b}\}\Big)=\int\prod_{\alpha}dq^{\alpha}\rho\,\Big(\{q^a,q^{\alpha}\},\{q^{\prime b},q^{\alpha}\}\Big).
\end{eqnarray}
Thus, the matrix $W$ can split into four blocks as,
\[ (W)_{AB} = \left(\begin{array}{ccc}
A_{ab} & B_{a\beta} \\
B^T_{\alpha b} & D_{\alpha \beta}\end{array} \right).\]
Consequently, the reduced density matrix is rewritten as,
\begin{align}
\rho_{red}(\{q^{a}\},\{q^{\prime b}\})=\left[\mathrm{det}\frac{M}{\pi}\right]^{1/2}e^{-\frac{1}{2}M_{ab}(q^{a}q^{b} +q'^{a}q'^{b})}e^{-\frac{1}{4}N_{ab}(q-q')^a (q-q')^b}
\label{rdm}
\end{align}
where
\begin{align}
M_{ab}=(A-BD^{-1}B^{T})_{ab} \qquad\text{and}\qquad N_{ab}=(B^TA^{-1}B)_{ab}.
\end{align}
The system can be diagonalized by the unitary matrix $U$ and the transformations
\begin{eqnarray}
q^{a}\rightarrow \tilde{q}^a=(UM^{1/2})^a_bq^{b}.
\end{eqnarray}
Thus, the density matrix can be further reduced into \cite{Bombelli},
\begin{align}
\rho_{red}(\{q^{a}\},\{q^{b}\})&=\Pi_{n}\left[\pi^{-1/2}\exp\left(-\frac{1}{2}(q_n q^n+q^{\prime}_n q^{\prime n}\right.\right.
-\left.\left.\frac{1}{4}\lambda_i (q-q^{\prime})_n (q-q^{\prime})^n)\right)\right],
\end{align}
where $\lambda_i$  are the eigenvalues of the matrix $\Lambda^a_b=(M^{-1})^{ac}N_{cb}$. Thus, the entanglement entropy of the system can be calculated by the relation \eqref{entanglemententropy} with
\begin{equation}
S_n=-\sum_{i}\ln(\lambda_i^{1/2}/2)+(1+\lambda_i)^{1/2}\ln(1+\lambda_i^{-1}+\lambda_i^{-1/2}).
\end{equation}

\begin{acknowledgments}
We thank Mahdis Ghodrati for her helpful suggestion on this manuscript. This work was supported by the Natural Science Foundation of China under Grant No.11705161 and Natural Science Foundation of Jiangsu Province under Grant No.BK20170481.
\end{acknowledgments}

\end{document}